\font\subtit=cmr12
\font\name=cmr8

\def\plb#1#2#3#4{#1, {\it Phys. Lett.} {\bf {#2}}B (#3), #4}
\def\npb#1#2#3#4{#1, {\it Nucl. Phys.} {\bf B{#2}} (#3), #4}

\def\cmp#1#2#3#4{#1, {\it Comm. Math. Phys.} {\bf {#2}} (#3), #4}

\def\ijmpa#1#2#3#4{#1, {\it Int. Jour. Mod. Phys.} {\bf A{#2}} (#3), #4}

\def\hpa#1#2#3#4{#1, {\it Helv. Phys. Acta} {\bf {#2}} (#3), #4}

\def\jgp#1#2#3#4{#1, {\it Journal Geom. Phys.} {\bf {#2}} (#3), #4}

\input harvmac
\null
{
\nopagenumbers
\hskip 8.2 cm {Preprint PAR-LPTHE 97-42}
\vskip 3truecm
\centerline{\subtit
OPERATOR FORMALISM FOR BOSONIC BETA-GAMMA FIELDS
}
\centerline{\subtit ON GENERAL ALGEBRAIC CURVES
}
\vskip 1truecm
\centerline{F{\name RANCO} F{\name ERRARI}$^{a}$ and J{\name AN} T.
S{\name OBCZYK}$^b$}
\smallskip $^a${\it
LPTHE, Universit\'e Pierre and Maria Curie--Paris VI and
Universit\'e Denis Diderot--Paris VII, Boite 126, Tour 16, 1$^{er}$ \'etage,
4 place Jussieu, F-75252 Paris Cedex 05, France, E-mail:
fferrari@lpthe.jussieu.fr
}
\smallskip $^b${\it
Institute for Theoretical Physics, Wroc\l aw
University, pl.  Maxa Borna 9, 50205 Wroc\l aw, Poland, E-mail:
jsobczyk@proton.ift.uni.wroc.pl}
\smallskip
\vskip 3cm
\centerline{ABSTRACT}
{\narrower \abstractfont
An operator formalism for bosonic $\beta -\gamma$ system on
arbitrary algebraic curves is introduced.
The classical degrees of freedom are identified and their
commutation relations are postulated. The explicit
realization of the algebra formed by the fields is given in the Hilbert space
equipped with a bilinear form.
The construction is based on the
"gaussian" representation for $\beta -\gamma$ system on the complex
sphere [Alvarez-Gaum\' e et al, Nucl. Phys. {\bf B 311} (1988) 333].
Detailed computations are provided for 2
and 4 point correlation functions.
} 
\Date{September 1997}}
\pageno=1
\vfill\eject
\newsec {INTRODUCTION}
\vskip 1cm
Since the middle of 1980s a big progress has been done in understanding
the structure of $D=2$ conformal field theories (CFT's)
\ref\cft{P. Di Francesco, P. Mathieu and D. S\' en\' echal, {\it Conformal
Field Theory}, Springer Verlag, 1996.}.
Most of that work
was motivated by string theory, but CFT's have also been
actively investigated because of their
role in describing phase transition phenomena. In string theory
it is necessary to consider
CFT's defined on topologically
nontrivial $D=2$ manifolds, in particular on closed higher genus Riemann
surfaces (RS) of genus $g$
\ref\pol{\plb{A. M.  Polyakov}{103}{1981}{207, 211}.
}.
The properties of CFT's on the torus are now relatively well
known and,
at least in the context of string theory, it turns out that the class
of acceptable models are those which are modular invariant
\ref\torus{
\npb{J. Cardy}{270[FS16]}{1986}{186}.
}.

There have been many efforts to give a satisfactory
description of CFT's also on higher genus RS, for instance
by means of an operator
formalism
\ref\altern{
\cmp{L. Bonora, A. Lugo, M. Matone and J. Russo}
{123}{1989}{329};
\plb{C. Vafa}{190}{1987}{47};
\npb{L. Alvarez-Gaum{\' e}, C. Gomez, G. Moore and C. Vafa}{303}{1988}
{455};
\plb{A. M. Semikhatov}{212}{1988}{357}.
}.
Usually, an operator formalism of such kind
makes it possible to introduce the notions of vacuum,
creation and annihilation operators,
normal ordering etc.
In this way, theories defined on curved space--times
resemble ordinary quantum field theories in 
flat space.
A valid operator formalism for string theory application should
be able to reproduce
the relevant correlation functions and their
analytical properties. Usually these properties
(the structure of zeros and poles) are strong enough
to determine these correlators up to an overall constant
\ref\vv{\npb{E. Verlinde and H.  Verlinde}{288}{1987}{357}},
\ref\raina{\hpa{A. K. Raina}{63}{1990}{694};
{\it Comm. Math. Phys.}, {\bf 140} (1991), 373.
}.

One version of  operator formalism has been developed for the $b-c$ systems
in
\ref\zn{\ijmpa{F. Ferrari and J. Sobczyk}{11}{1996}{2213}.},
\ref\gen{\jgp{F. Ferrari and J. Sobczyk}{19}{1996}{287}.}. In this
paper we would like to extend it in such a way that also the bosonic
$\beta -\gamma$ systems can effectively be treated.

These systems are in fact
interesting examples of CFT's.
They have been extensively studied in the second half of
1980s after they
turned out to be a necessary ingredient (as Faddeev-Popov
superghosts) in the
perturbative approach to strings
\ref\fms{
\npb{D. Friedan, E. Martinec and S.
Shenker}{271}{1986}{93}.
}\nref\atisen{\npb{J. Atick and A. Sen}{293}{1987}{317}.}
\nref\vvb{
\plb{E. Verlinde and H. Verlinde} {192}{1987}{95}.
}--\ref\bega{
\plb{A. Losev}{226}{1989}{62};
\plb{O. Lechtenfeld}{232}{1989}{193};
\plb{P. di Vecchia}{248}{1990}{329};
\plb{U. Carow-Watamura, Z. F. Ezawa, K. Harada, A. Tezuku and S. Watamura}
{227}{1989}{73}.
}. A treatment on hyperelliptic curves, mainly in connection with
the computation
of amplitudes in superstring theory at two loops,
can for instance be found in refs. 
\ref\hyperelliptic{\npb{D. Montano}{297}{1988}{125}; E. Gava,
{\it Conformal Fields on Complax Curves}, Preprint LPTENS-88/01;
\plb{E. Gava, R. Iengo and G.Sotkov}{207}{1988}, 283; \npb{D. Lebedev
and A. Morozov}{302}{1988}{163}; \plb{R. Iengo and C-J Zhu}{212}{1988}{313}.}.
$\beta-\gamma$ systems have also been studied in the context
of the Wess-Zumino-Witten model
\ref\wzw{\ijmpa{A. Gerasimov, A. Marshakov, A. Morozov et al.}{5}{1990}{2495}.
}. 

The analysis of the $\beta -\gamma$ systems is in many respects analogous to
that of the $b-c$ systems. In fact, the only difference lies in the
statistics, since the
$\beta -\gamma$ fields commute while the $b-c$ fields anticommute.
Following \fms,  many formulas characterizing both systems can be written in
common. If by {\bf b} (and {\bf c}) one denotes either the $b$ or $\beta$
($c$ or $\gamma$) fields with conformal dimension $\lambda$ (and $1-\lambda$),
then both theories are defined by the action:

\eqn\act{
S = {1\over \pi}\int d^2z ({\bf b}\bar \partial {\bf c}).
}

where $\bar \partial\equiv {\partial\over \partial\bar z}$.
It follows from \act\ that the OPE's for the fields are

\eqn\opbc{
{\bf c}(z){\bf b}(w) \sim {1\over z-w},\qquad
{\bf b}(z){\bf c}(w) \sim {\epsilon\over z-w}
}

where $\epsilon =1$ for $b-c$ systems and $\epsilon = -1$ for
$\beta -\gamma$ systems. The energy-momentum tensor, determined by the
dependence of \act\ on the metric (in order to derive it one has to
write the action in the covariant form) carries no dependence on
$\epsilon$:

\eqn\enmom{
T= -\lambda {\bf b}\partial {\bf c} + (1 -\lambda) (\partial {\bf b})
{\bf c}.
}

The OPE's of the elementary fields with $T$ can be described with the
same formula both in the fermionic and bosonic case,
as they are determined only
by conformal dimensions:

\eqn\optb{
T(z){\bf b}(w) \sim {\lambda\over (z-w)^2} {\bf b}(w) + {1\over z-w} \partial
{\bf b}(w) + ...
}

\eqn\optc{
T(z){\bf c}(w) \sim {1- \lambda\over (z-w)^2} {\bf c}(w) + {1\over z-w} \partial
{\bf c}(w) + ...
}

The OPE with two energy-momentum tensors depends on
$\epsilon$ since in calculating it one has to (anti)-commute elementary fields:

\eqn\optt{
T(z)T(w) \sim {-\epsilon (6\lambda^2-6\lambda +1)\over (z-w)^4}
+ {2 T(w)\over (z-w)^2} + {\partial T(w)\over z-w} + ...
}

The correlation functions of the $\beta -\gamma$ systems
were intensively studied
at the end of the last decade \vvb, \bega.

In string theory computations, one would like to compute
the correlators also within the bosonized version of the theory.
It turns out that in
addition to the scalar fields with quadratic action one has to
introduce also auxiliary fermionic fields $\eta -\xi$ of conformal dimensions
$1$ and $0$. One gets an infinite number of inequivalent representations
($q$ vacua in \fms ) of the basic algebra
in a Hilbert space.

For $\lambda\geq 1$ and $g\geq 2$ only the 
zero modes in the {\bf b} fields are
present. In both bosonic and fermionic cases one has to insert
delta functions for any zero mode
in order to obtain a reasonable
correlator. Without these insertions one gets vanishing or diverging amplitudes
in the case of fermionic or bosonic systems respectively.
In the path integral
approach this follows from the
properties of the integrals:

\eqn\propint{\int_{\bf R} dx \rightarrow \infty,\qquad \int d\zeta = 0}

where $\zeta$ is an element of a Grassmann algebra satisfying $\zeta^2=0$.
In the fermionic case $\delta(\zeta )=\zeta$, so that any insertion in the
amplitudes
of a delta function $\delta(b)$
is equivalent to an insertion of a $b$ field.
In the bosonic case one has instead to
make sense of true delta functions (or step functions) containing
field operators in their argument.

The manipulations with delta function operators become straigthforward in
the "gaussian" representation for $\beta$ and $\gamma$ fields
\ref\ag{
\npb{L. Alvarez-Gaum\' e, C. Gomez, P. Nelson, G. Sierra and C. Vafa}
{311}{1988}{333}.
}.
In that representation
one sees that inequivalent vacua can be chosen and that they
cannot be
interchanged under the action of a finite number of elementary excitations.
Each vacuum state is determined by a choice of signs in a infinite
product of gaussian (or, strictly speaking "fresnelian") functions
$\exp (\pm {i\over 2} x^2)$.
The general
property of delta-like operators is that they
can map one vacuum state into another
\ag . Some operations in the gaussian representation may seem formal, e.g.
formally infinite normalization factors have to be included. However,
there is
always a unique procedure to handle with these factors in
order to obtain finite results.

The gaussian representation makes it possible to define an operator formalism
for the  $\beta -\gamma$ systems following the ideas of \gen, where  the
case of the $b-c$
systems has been discussed. In \gen, the classical degrees of freedom are
identified and the simplest possible anticommutation relations are postulated.
Some of the elementary oscillators are defined as annihilation or
creation operators. The remaining excitations correspond to the
zero modes of the model. It has been demonstrated that these
definitions enable the calculation of the correlation functions of the theory.
The construction is not restricted to any
particular class of RS - it can be done on arbitrary algebraic curve.

The aim of this paper is to present in detail the construction
of an analogous operator
formalism on RS for $\beta -\gamma$ systems.
The outcome of our
computations are correlation functions with all the necessary
analytic properties.
%
%


The operator formalism for the $\beta -\gamma$ systems
exhibits some similarities and also some interesting
differences with respect
to
the operator formalism found in the case of the $b-c$ fields in \gen.
The starting point is the same, one chooses the same
classical degrees of freedom and postulates the simplest possible
commutation (instead of anticommutation)
relations. One decomposes the elementary excitations
into annihilation, creation and zero mode sectors. A fundamental
property of the normal ordering is that the difference between an ordered and
an
unordered
product of two fields is equal to
the Weierstrass kernel $K_{\lambda}(z,w)$,
which is one of the basic buiding blocks in the construction of
the correlation
functions.
$K_{\lambda}(z,w)$ is a $\lambda$ differential in $z$, a $1-\lambda$
differential
in $w$ and its only finite
singularity in $z$ is a single pole at $z=w$ (provided
$z$ and $w$ parametrize points on the same branch of the RS).
In our investigations
we need to introduce
multivalued fields on the complex sphere.
In this sense one can think that new classes of CFT's containing
multivalued models are defined on the sphere
\zn.
The computation of the correlation functions
for $\beta -\gamma$ systems is quite different
from the case of the $b-c$ systems. In the bosonic
case many nontrivial manipulations with Fresnel integrals are
required before reaching
the final result. This will be shown in detail in Chapters
2-4.

The experience from the $\beta -\gamma$ systems can hopefully be
used in dealing with more complicated and important systems, such for instance
the bosonic theories with quadratic action. These theories provide the basis
of the Coulomb gas formulation of minimal models. Therefore, the
construction of an operator formalism for the $\beta-\gamma$
systems can shed some light in the treatment of the
minimal models on arbitrary RS represented as algebraic curve.
Until now,
only the case of hyperellliptic and $Z_N$ symmetric curves
have been studied in this
context
\ref\crnk{\plb{C. Crnkovic, G.M. Sotkov and M. Stanishkov}{220}{1989}{397}.},
\ref\ae{S. A. Apikyan and C. J. Efthimiou, {\it Minimal Models of CFT's
on $Z_N$ Surfaces}, hep-th 9610051.}.

\vskip 1cm
\newsec{$\lambda =-1$ COMPUTATIONS ON THE SPHERE}
\vskip 1cm
To better explain the ideas of the
calculations, it is useful to start with the
simpler situation of the complex sphere. We choose in this Chapter $g=0$
and
$\lambda =-1$ in order to have $\beta$ rather then $\gamma$ field zero
modes.

The classical degrees of freedom are identified as

\eqn\xbexp{\beta (z) = \sum_{j=-\infty}^{+\infty}
\beta_{j} z^{-j+1}dz^{\lambda} }

\eqn\xgexp{\gamma (z) = \sum_{k=-\infty}^{+\infty}
\gamma_{k} z^{-k-2}dz^{1-\lambda} }

In quantum theory one postulates the following commutation relations:

\eqn\xbasrel{ [\gamma_{k}, \beta _{j}] = \delta_{j+k,0} .}

The normal ordering is defined by assigning to the elementary excitations
$\beta_{j}$ and $\gamma_{k}$
the property that they are creation or annihilation operators.
Annihilation operators  correspond to $j\geq 2$ and $k\geq -1$.
Creation operators are those with $j\leq -2$ and $k\leq
-2$.
The remaining $\beta_{j}$ operators ($j=0, \pm 1$) correspond to the
three zero
modes (holomorphic vector fields).

The above assignement implies:

\eqn\xnoord{\gamma (z)\beta (w) = :\gamma (z)\beta (w): + {1\over z-w}. }

which should be understood as an identity in the radial ordered correlation
functions.

Another important identity is
(in $\exp \big(ip\beta (z)\big)$
the introduction of normal ordering makes no
modification):

\eqn\xnoorexp{\gamma(z) {\rm e}^{ip\beta (w)} =
:\gamma(z) {\rm e}^{ip\beta (w)}: + {ip\over z-w} {\rm e}^{ip\beta
(w)}. }

Let us start calculating propagator of the theory.
It is known from the general discussion
and it will also be very clearly seen in the course of the
computations that the propagator
requires 3
(the number of $\beta$ field zero modes) insertions of
delta functions of $\beta$ fields.

Using the Wick theorem one calculates:

$$<0|\gamma (z) \beta (w) \exp \left(
i\sum_{r=1}^{3} p_r\beta (w_r)\right) |0> =$$
$$={1\over z-w} <0|
\exp \left( i\sum_{r=1}^3\sum_{j=-\infty}^{+\infty}
p_r\beta_{j}w_r^{-j+1} \right)  |0> +$$
\eqn\propa{ +\sum_{r=1}^3 {ip_r\over z-w_r}
<0|\beta (w)  \exp \left( i\sum_{t=1}^3\sum_{j=-\infty}^{+\infty}
p_t\beta_{j}w_t^{-j+1}
\right)  |0>.}

Finally, the integration over $\prod_{r=1}^3{dp_r\over 2\pi}$ has to be
performed.

The computations will be done in the "gaussian" representation for the
$\beta -\gamma$ systems \ag .
The basic definitions of
that construction are summarized
in Appendix A.

In order to evaluate \propa\ one calculates

\eqn\mea{ \exp \left( i\sum_{n=-\infty}^{+\infty} A_n\beta_n\right) |0> }

with the state $|0>$ taken to be the state $\Phi^{(-1)}$ defined in (A.8)

\eqn\vl{
|0> = \Phi^{(-1)} = \exp {i\over 2} \left(
\sum_{n\geq -1} x_n^2 - \sum_{n<-1}x_n^2\right).
}

A concrete realization of the operators $\beta_n$ is given by (see (A.2)): 

\eqn\vbr{ \beta_n = - {i\over \sqrt{2}} (x_{-n} -i {\partial \over
\partial x_{-n}})}

The operators $\beta_n$ involve only $x_{-n}$
and acting on
$\exp \left( -{i\over 2} x_{-n}^2\right)$ give $0$.

It follows that

\eqn\meca{ \exp \left( i\sum_{n=-\infty}^{+\infty}
 A_n\beta_n\right) |0> =
\exp \left( -{i\over 2}\sum_{m<-1} x_m^2\right)
\exp \left( i\sum_{n\leq 1} A_n\beta_n\right)
\exp \left( {i\over 2}\sum_{m\geq -1} x_m^2\right). }

Repeating steps similar to those presented
in the end of Appendix A one finds

$$\exp \left( -{i\over 2}\sum_{m<-1} x_m^2\right)
\exp \left( i\sum_{n\leq 1} A_n\beta_n\right)
\exp \left( {i\over 2}\sum_{m\leq 1} x_{-m}^2\right) =$$
\eqn\funa{
=\exp \left( -{i\over 2}\sum_{m<-1} x_m^2\right)
\exp \left( -{i\over 2}\sum_{n\leq 1} \big( (A_n + i\sqrt{2}x_{-n})^2
+  x_{-n}^2 \big) \right) }

To define a bilinear form we introduce the right vacuum:

\eqn\bra{ <0| = \exp  {i\over 2}\left( \sum_{m>1} x_m^2
- \sum_{m\leq 1} x_m^2 \right) . }

A proper normalization of scalar product is included in
a (formally infinite) factor $V$:

$$<0| \prod_{s=1}^3  \delta (\beta (z_s) |0>
= {1\over V}
\int \prod_{s=1}^3dp_s
\int \prod_{k=-\infty}^{+\infty} dx_k
\exp \left( -i \sum_{m<1}x_m^2\right)\times$$

\eqn\prko{
\times\exp \left(
-{i\over 2}\sum_{m>1}
\left( A_{-m} + i\sqrt{2} x_m\right)^2\right)
\exp \left( - {i\over 2}
\sum_{m=0,\pm 1}\big( (A_m + i\sqrt{2} x_{-m})^2 + 2 x_{-m}^2)\big)
\right) }

for $A_m = \sum_{s=1}^3 p_s z_s^{1-m}$. It will soon become clear that
$V$ should be defined as

$$V= \Big(\prod_{m<1} \int \exp \left( -i x_m^2\right) dx_m\Big)
\Big(\prod_{n>1}\int \exp \left( i x_n^2\right) dx_n\Big)\times$$

\eqn\vnorm{
\times
\Big(\prod_{k=0,\pm 1}\int \exp \left( -{i\over 2} p_k^2\right) dp_k\Big)
\Big(\prod_{l=0,\pm 1}\int \exp \left( -i x_l^2\right) dx_l\Big).
}

It is most convenient to integrate first
over $\prod_mdx_m$ for $m<1$ and then for $m>1$. Both integrations
are trivial and contribute only to overall normalization.
Next one should
integrate over $\prod_{s=1}^3 dp_s$ after a convenient
change of variables that will produce an appropriate Jacobian.
The trivial integration over
$\prod_{m=0,\pm 1}dx_m$ will be performed later.

The above mentioned change of variables is:

\eqn\cva{p_j\rightarrow A_m,\qquad j=1,2,3,\qquad m=0,\pm 1.}

\eqn\cvb{\int \prod_{j=1}^3 dp_j = \int \prod_{m=-1}^1 dA_m \det \left(
{\partial A_n\over \partial p_k}\right)^{-1}
}

where

\eqn\cvc{ \det \left( {\partial A_n\over\partial
p_k}\right)
=(w_1-w_2)(w_1-w_3)(w_2-w_3).}

The integration over $dp_s$ and $dx_n$ cancel precisely the factor
$V$ and
the final result is

\eqn\abfin{ <0| \prod_{s=1}^3  \delta (\beta (w_s) |0> =
{1\over (w_1-w_2)(w_1-w_3)(w_2-w_3)}. }

In the computation of the second term in \propa\ it is sufficient to
concentrate ourselves on the
most important point without
repeating many of the above arguments.

$$\beta (w) \exp \left( i\sum_{n=-\infty}^{+\infty}
A_n\beta_n\right) |0> =$$

$$= -{i\over\sqrt{2}}\sum_{j=-\infty}^{+\infty}
w^{-j+1} \left( x_{-j} -i{\partial\over\partial
x_{-j}}\right)
\exp \left( -{i\over 2}\sum_{m<-1} x_m^2\right)\times$$

$$\times\exp \left( -{i\over 2}\sum_{n\leq 1} ((A_n + i\sqrt{2}x_{-n})^2
+  x_{-n}^2 ) \right) =$$

$$= - \exp \left( -{i\over 2}\sum_{m<-1} x_m^2\right)
\sum_{j\leq 1} w^{-j+1} \left( A_j + i\sqrt{2} x_{-j} \right)\times$$

\eqn\abc{
\times\exp \left( -{i\over 2}\sum_{n\leq 1} \big( (A_n + i\sqrt{2}x_{-n})^2
+  x_{-n}^2 \big) \right) }

With the above definition of the right vacuum $<0|$ one gets

$$<0| \beta (w) \exp \left( i\sum_{n-\infty}^{+\infty}
A_n\beta_n\right) |0> =
- {1\over V} \int \prod_{s=1}^3 dp_s
\int \prod_{k=-\infty}^{+\infty}
dx_k
\sum_{j\leq 1} w^{-j+1} \left( A_j + i\sqrt{2} x_{-j} \right)$$

$$\times\exp \left( -i \sum_{m<1}x_m^2\right)
\exp \left(
-{i\over 2}\sum_{m>1}\left( A_{-m} + i\sqrt{2} x_m\right)^2\right)\times$$

\eqn\abd{
\times\exp \left( - {i\over 2}
\sum_{m=0,\pm 1}\big( (A_m + i\sqrt{2} x_{-m})^2 + 2 x_{-m}^2\big)\right) }

As before, it is most convenient to perform first the integration over
$\prod_{m<-1} dx_m$ and then over $\prod_{m>1} dx_m$.
The sum over $j\leq 1$ reduces to three terms with $j=0,\pm 1$.
All the other terms vanish due to integration over antisymmetric
function in $x_n$ $(n>1)$.

Having in mind future generalizations, the following notation is useful:

\eqn\chva{ A_j = \sum_{k=1}^3 R_{jk} p_k. }

Thus, in computating the second term in \propa\ one is left with the evaluation
of:

$$-i {1\over V'} \int\prod_{j=0,\pm 1} dA_j\int \prod_{n=0,\pm 1}dx_n
\det \left(
{\partial A_m\over \partial p_r}\right)^{-1}\times$$

\eqn\cel{
\times\sum_{s=1}^3\sum_{j,k=-1}^{1} {R^{-1}_{sk} A_k\over z-w_s} w^{1-j}
(A_j + i \sqrt{2} x_{-j})
\exp\left( -{i\over 2} \sum_{l=-1}^1 \left( (A_l + i\sqrt{2} x_{-l})^2 +
2 x_l^2 \right)\right). }

with

\eqn\nordwa{
V'= \Big( \prod_{k=0,\pm 1}\int \exp \left( -{i\over 2} p_k^2\right) dp_k\Big)
\Big( \prod_{l=0,\pm 1}\int \exp \left( -i x_l^2\right) dx_l\Big) .}

It is useful to perform still another change of variables:

\eqn\chvab{ \tilde A_j = A_j + i\sqrt{2} x_{-j} }

Thus one can rewrite \cel\ as

$$-i {1\over V'} \int\prod_{j=0,\pm 1} d\tilde A_j
\int\prod_{n=0,\pm 1}dx_n
\Big(\det \left(
{\partial \tilde A_m\over \partial p_r}
\right)\Big)^{-1}\times$$

\eqn\cela{
\times\sum_{s=1}^3\sum_{j,k=-1}^1
{R^{-1}_{sk} (\tilde A_k -i\sqrt{2} x_{-k})\over z-w_s}
w^{1-j}
\tilde A_j \exp \left(
-{i\over 2} \sum_{l=-1}^1 \left( \tilde A_l^2 + 2 x_l^2 \right)\right) }

The integration over $\prod_{j=-1}^1
d\tilde A_j$ gives nonzero result only for
$k=j$. Terms with $i\sqrt{2}x_{-k}$ drop out (they give rise to integral
linear in $\tilde A_{j}$). The
integration over $\prod_{l=-1}^1 dx_l$
is then trivial and one obtains:

\eqn\celb{ -
\sum_{s=1}^3\sum_{j=-1}^1
{R^{-1}_{sj}\over z-w_s} w^{1-j}.
}

In order to get the numerical factor in
\celb\ one has to use

\eqn\nf{\int_{\bf R} A^2 {\rm e}^{-{i\over 2}A^2}dA =
-i \int_{\bf R} {\rm e}^{-{i\over 2}A^2}dA
}

The matrix $R^{-1}$ is of the form:

\eqn\rmat{ R^{-1} = {1\over (w_1-w_2)(w_1-w_3)(w_2-w_3)}
\pmatrix{ w_2-w_3&w_3^2-w_2^2&w_2w_3(w_2-w_3)\cr
w_3-w_1&w_2^2-w_3^2&w_1w_3(w_3-w_1)\cr
w_1-w_2&w_w^2-w_1^2&w_1w_2(w_1-w_2)} }

and the final result is (one has to include $\det {\partial\tilde
A\over \partial p}$):

$$<0|\gamma (z) \beta (w) \prod_{s=1}^3 \delta (\beta (w_s)) |0> =$$

$$= {1\over (w_1-w_2)(w_1-w_3)(w_2-w_3)} \Big( {1\over z-w}
 - {(w-w_2)(w-w_3)\over (z-w_1)(w_1-w_2)(w_1-w_3)} +$$

$$+ {(w-w_1)(w-w_3)\over (z-w_2)(w_1-w_2)(w_2-w_3)}
- {(w-w_1)(w-w_2)\over (z-w_3)(w_1-w_3)(w_2-w_3)} \Big) =$$


\eqn\absfin{
= {1\over z-w}\prod_{s=1}^3{w-w_s\over z-w_s}\prod_{m<n}(w_m-w_n)^{-1}
}

It is clear that the above expression has all the
analytical properties that the propagator
of the bosonic $\beta -\gamma$ systems should possess.

\vskip 1cm
\newsec{THE PROPAGATOR ON ARBITRARY ALGEBRAIC CURVES}
\vskip 1cm
In this Chapter the $\beta -\gamma$ system with $\lambda\geq 1$
given by \act\
is defined on arbitrary RS represented
by means of algebraic equation

\eqn\aleq{ F(z,y) = y^NP_N(z) + y^{N-1}P_{N-1}(z) + ... + P_0(z) = 0 }

where $P_j(z)$ are polynomials in $z$. $y$ can be viewed either as a
singlevalued
functions on the RS or
as multivalued function on the complex sphere.
The monodromy
properties of $y(z)$ then define the RS.

The classical degrees of freedom are identified as

\eqn\bexp{\beta (z) = \sum_{s=0}^{N-1}\sum_{j=-\infty}^{+\infty}
\beta_{s,j} z^{-j-\lambda} f_s(z)}
\eqn\gexp{\gamma (z) = \sum_{p=0}^{N-1}\sum_{k=-\infty}^{+\infty}
\gamma_{p,k} z^{-k+\lambda -1} \phi_p(z)}

where

\eqn\dfk{
f_k(z) = {y^{N-1-k}(z)dz^{\lambda}\over (F_y(z, y(z)))^{\lambda}}
}

and

\eqn\dfik{
\phi_k(z) = {dz^{1-\lambda}\over (F_y(z, y(z)))^{1-\lambda}}
\Big( y^k(z) + y^{k-1}(z)P_{N-1}(z) + ... + P_{N-k}(z)\Big) .
}

The basic commutation relations are:

\eqn\basrel{ [\gamma_{p,k}, \beta _{s,j}] = \delta_{s,p}\delta_{j+k,0} .}

The price paid for the simple commutation relations \basrel\ is the
complicated definition of basis \dfik .

The
normal ordering is introduced by assigning to ethe lementary excitations
$\beta_{s,j}$ and $\gamma_{s,k}$
the property that they are creation or annihilation operators.
Annihilation operators are those with
$j\geq 1-\lambda$ and $k\geq \lambda$ and
creation operators with $j\leq-\lambda - M_s$ and $k\leq
\lambda -1$ ($M_s$ is a number of zero modes in the $s$ sector of the
theory). The remaining operators $\beta_{s,j}$ with
$-\lambda-M_s<j<1-\lambda$ correspond to zero modes.

From the above assignement it follows that:

\eqn\noord{\gamma (z)\beta (w) = :\gamma (z)\beta (w): + {1\over z-w}
\sum_{n=0}^{N-1} f_n(z)\phi_n(w) = :\gamma (z)\beta (w): + K_{\lambda}
(z,w). }

It is an identity in radial ordered correlation
functions.
$K_{\lambda}(z,w)$ is the Weierstrass kernel (see Introduction).

A simple implication of \noord\ is

\eqn\noorexp{\gamma(z) {\rm e}^{ip\beta (w)} =
:\gamma(z) {\rm e}^{ip\beta (w)}: + ip K_{\lambda}(z,w) {\rm e}^{ip\beta
(w)}. }

We wish to calculate the propagator for $\beta -\gamma$ system.
In order to have finite correlators one has to insert a number
$M_{tot}$ of delta functions in the $\beta$ fields, where $M_{tot}$ is the
total number of $\beta$ zero modes:

\eqn\zmtot{
M_{\rm tot} = \sum_{s=0}^{N-1} M_s = (g-1)(2\lambda -1)
}

The Wick theorem implies

$$<0|\gamma (z) \beta (w) \exp \left(
i\sum_{r=1}^{M_{\rm tot}} p_r\beta (w_r)\right) |0> =$$
$$=K_{\lambda} (z,w) <0|
\exp \left( i\sum_{r=1}^{M_{\rm tot}}\sum_{s=0}^{N-1}\sum_{j=-\infty}^{+\infty}
p_r\beta_{s,j}w_r^{-j-\lambda}
f_s(w_r)\right)  |0> +$$
\eqn\propb{ +\sum_{r=1}^{M_{\rm tot}} ip_rK_{\lambda} (z, w_r)
<0|\beta (w)  \exp \left( i
\sum_{t=1}^{M_{\rm tot}}\sum_{s=0}^{N-1}\sum_{j=-\infty}^{+\infty}
p_t\beta_{s,j}w_t^{-j-\lambda}
f_s(w_t)\right)  |0>.}

The propagator is obtained after
integration over
$\prod_{r=1}^{M_{\rm tot}}{dp_r\over 2\pi}$.

As in Chapter 2, the
computations will be performed in the "gaussian" representation for the
$\beta -\gamma$ system \ag. However,
complications can be expected on a Riemann surface
due to appearence of $M_s$ zero modes in the sector $s$ of the theory.

A first attempt to accomodate these zero modes is
to modify the way in which the $\beta_{s,j}$ and $\gamma_{p,k}$
excitations are realized:

\eqn\tbrep{\beta_{s,j} = - {i\over \sqrt{2}} (x_{s,-j-\nu_s} -i
{\partial \over \partial x_{s,-j-\nu_s}})}

\eqn\tgrep{\gamma_{p,n} =  {1\over \sqrt{2}} (x_{p,n+\nu_p} +i
{\partial \over \partial x_{p,n+\nu_p}})}

where $\nu_s$ is to be related to
the number $M_s$ of zero modes in the $s$ sector.

Let us suppose that the vacuum state is taken as:

\eqn\tvacdefc{ \Phi^{\bar\kappa}(x) = \exp \left( {i\over 2} \sum_{s=0}^{N-1}
\left(
\sum_{n\geq \kappa_s} x_{s,n}^2 - \sum_{n<\kappa_s}x_{s,n}^2\right)\right) }

where $\kappa_s$
is again chosen in an appropriate way. Unfortunately, it turns out that
nothing can be achieved in this way. The conditions that $\beta_{s,j}$ and
$\gamma_{s,k}$ are annihilation operators only for $j\geq 1-\lambda$,
$k\geq\lambda$ imply in fact that

\eqn\npw{ \kappa_s = \lambda,\qquad \nu_s = 0.}

We conclude that the only way to accomodate the generic structure of zero
modes is to redefine the scalar product in an appropriate way. We shall come
back to this point shortly.

In calculating \propb\ one follows most of steps described in Chapter 2.
The vacuum state is that introduced in \tvacdefc\ with $\kappa_s=\lambda$.
One finds

$$\exp \left( i\sum_{s=0}^{N-1}\sum_{j=-\infty}^{+\infty}
A_{s,j}\beta_{s,j}\right) |0> = \exp \left( i\sum_{s=0}^{N-1}
\sum_{j\leq\lambda} A_{s,j}\beta_{s,j}\right) |0> =$$

\eqn\asia{ = \exp \left( -{i\over 2}\sum_{s=0}^{N-1}\left(
\sum_{n<\lambda} x_{s,n}^2
+ \sum_{j\leq -\lambda}\big(  (A_{s,j} +
i\sqrt{2}x_{s,-j})^2 + x_{s,-j}^2\big) \right)\right)
}

with

\eqn\asj{
A_{s,j} = \sum_{r=1}^{M_{\rm tot}} p_rw_r^{-j-\lambda}f_s(w_r)
}

The correct definition of $<0|$ is

\eqn\bra{ <0| = \exp \left( {i\over 2} \sum_{s=0}^{N-1}
\left( \sum_{n\geq\lambda +M_s}x_{s,n}^2 - \sum_{n<\lambda +M_s}
x_{s,n}^2\right)\right) }

A possible way of thinking about $<0|$ is that under the conjugation

\eqn\xcon{ x_{s,n}^{\dagger} = x_{s,-n+2\lambda +M_s-1.}
}

The conjugation on elementary excitations can be deduced from \tbrep\
and \tgrep\ to be

\eqn\scon{ \beta_{s,n}^{\dagger} = - \beta_{s,-n-2\lambda-M_s+1},\qquad
\gamma_{s,n}^{\dagger} = \gamma_{s,-n+2\lambda +M_s-1}.
}

This definition of conjugation
is analogous
to that obtained for the $b-c$ system on arbitrary RS
\ref\nowab{
F. Ferrari, J. T. Sobczyk, {\it Monodromy Properties of
Energy-Momentum Tensor on General Algebraic Curves}, Preprint PAR-LPTHE
97-38, hep-th/9709162
}.
For $g=0$ and $\lambda =-1$ (the situation
discussed in Chapter 2) $M_{\rm tot}=3$
and $\beta_n^{\dagger} = -\beta_{-n}$,
$\gamma_n^{\dagger} = \gamma_{-n}$.

The first term in \propb\ is

$${1\over {\bf V_{\lambda}}}\int
\prod_{s=0}^{N-1} \prod_{j=-\infty}^{+\infty}dx_{s,j} \int
\prod_{r=1}^{M_{\rm tot}}dp_r
\exp\left( {i\over 2}\sum_{s=0}^{N-1}\left( \sum_{n\geq\lambda +M_s}
x_{s,n}^2 - \sum_{n<\lambda + M_s} x_{s,n}^2\right)\right)\times$$

\eqn\raz{
\times\exp\left( -{i\over 2} \sum_{s=0}^{N-1}\left(
\sum_{n<\lambda}x_{s,n}^2 + \sum_{j\geq\lambda}\left( (A_{s,-j}
+i\sqrt{2}x_{s,j})^2 + x_{s,j}^2\right)\right)\right).
}

The normalization factor (formally infinite) ${\bf V_{\lambda}}$ is

$${\bf V_{\lambda}} = \Big( \int\prod_{s=0}^{N-1}\prod_{n<\lambda}\exp\left(
-ix_{s,n}^2\right) dx_{s,n}\Big)
\Big( \int\prod_{s=0}^{N-1}\prod_{n\geq\lambda +M_s}\exp\left(
ix_{s,n}^2\right) dx_{s,n}\Big)\times$$

\eqn\dan{
\times\Big(
\int\prod_{r=1}^{M_{\rm tot}}\exp\left( -{i\over 2}p_r^2\right) dp_j\Big)
\Big( \int\prod_{s=0}^{N-1}\prod_{n=\lambda}^{\lambda +M_s-1}
\exp\left( -ix_{s,n}^2\right) dx_{s,n}\Big)
}

One integrates first over
$\prod_{s=0}^{N-1}\prod_{n<\lambda}dx_{s,n}$ and then over
$\prod_{s=0}^{N-1}
\prod_{n\geq\lambda +M_s}dx_{s,n}$. What remains is

$${1\over {\bf V'_{\lambda}}} \int
\prod_{s=0}^{N-1}\prod_{n=\lambda}^{\lambda +M_s-1}dx_{s,n}
\int \prod_{r=1}^{M_{\rm tot}}\exp\left( -{i\over 2}p_r^2\right) dp_j$$

\eqn\fiol{
\times \exp\left( -{i\over 2}\sum_{s=0}^{N-1}\sum_{n=\lambda}^{\lambda
+M_s-1}\left( (A_{s,-n}+i\sqrt{2}x_{s,n})^2 + 2
x_{s,n}^2\right)\right).}

with

\eqn\dana{
{\bf V'_{\lambda}} =
\Big(\int\prod_{r=1}^{M_{\rm tot}}\exp\left( -{i\over 2}p_r^2\right) dp_j\Big)
\Big(\int\prod_{s=0}^{N-1}\prod_{n=\lambda}^{\lambda +M_s-1}
\exp\left( -ix_{s,n}^2\right) dx_{s,n}\Big).
}

The number of integrations to be done over $dx_{s,n}$ is equal $M_{\rm
tot}$. It is very convenient to introduce an index $L$ which runs over
all the allowed pairs $(s,n)$ $L=1, ..., M_{\rm tot}$. There is still
minor technical complications due to the fact that in \fiol\ expressions
with $x_{s,n}$ are combined with $A_{s,-n}$. The following definition will
be adopted: if the index $J$ denotes a pair $(s,n)$ then

\eqn\juz{
x_J\equiv x_{s,n},\qquad A_J\equiv A_{s,-n}.
}

A change of variables is to be done

\eqn\cvt{ p_j\longrightarrow A_{r,-k} =
\sum_{n=1}^{M_{\rm tot}} p_n w_n^{k-\lambda} f_r(w_n)
= \sum_{n=1}^{M_{\rm tot}} p_n \Omega_L (w_n) }

$\Omega_L(w)$ are $\beta$ field zero modes, i.e. holomorphic $\lambda$
differentials.
Introducing a similar notation as in Chapter 2 one can write

\eqn\cvtb{ A_L = \sum_{j=1}^{M_{\rm tot}} R_{Lj}p_j }

where

\eqn\cvtc{ R_{Lj} = \pmatrix {
\Omega_1(w_1)&\Omega_1(w_2)&...&\Omega_1(w_{M_{\rm tot}}) \cr
...&...&...&...\cr
\Omega_{M_{\rm tot}}(w_1)&\Omega_{M_{\rm tot}}(w_2)&...
&\Omega_{M_{\rm tot}}(w_{M_{\rm tot}}) } }

is a matrix of zero modes. The result is that only the Jacobian arising
due to the
change of variables \cvt\ gives nontrivial contribution. The remaining
integrations cancel the normalization factor ${\bf V'_{\lambda}}$ and the
first term in
\propb\ is

\eqn\tft{
{K_{\lambda} (z,w)\over \det \Omega_L(w_j)}. }

In order to calculate the next term in \propb\ one should follow the
procedure presented in Chapter 2 with the above explained
modifications due to presence of
$N$ sectors and $M_s$ zero modes in each of them. The
result is

\eqn\tftb{
- \det \big(\Omega_L(w_j)\big)
\sum_{s,N=1}^{M_{\rm tot}} K_{\lambda}(z, w_{s})
R_{sN}^{-1} \Omega_{N}(w).
}

The final result can be written in compact form

$$<0|\gamma (z) \beta (w) \prod_{r=1}^{M_{\rm tot}}
 \delta (\beta (w_r)) |0> =$$
\eqn\tfin{ = \det\pmatrix{
K_{\lambda}(z,w)&K_{\lambda}(z,w_1)&...&K_{\lambda}(z,w_{M_{\rm tot}})\cr
\Omega_1(w)&\Omega_1 (w_1)&...&\Omega_1(w_{M_{\rm tot}})\cr
...&...&...&...\cr
\Omega_{M_{\rm tot}} (w)&\Omega_{M_{\rm tot}} (w_1)
&...&\Omega_{M_{\rm tot}} (w_{M_{\rm tot}})}
(\det \Omega_L(w_j))^{-2}. }

It has all the necessary analytical properties of the propagator: poles
as
$z\rightarrow w$, $z\rightarrow w_r$, $w_{r_1}
\rightarrow w_{r_2}$ ($r_1\neq r_2)$ and zeroes as $w\rightarrow w_j$.

One can also define

$${<0|\gamma (z) \beta (w) \prod_{r=1}^{M_{\rm tot}}
\delta (\beta (w_r)) |0>
\over <0| \prod_{r=1}^{M_{\rm tot}} \delta (\beta (w_r)) |0> } =$$

\eqn\tfina{ = \det\pmatrix{
K_{\lambda}(z,w)&K_{\lambda}(z,w_1)&...&K_{\lambda}(z,w_{M_{\rm tot}})\cr
\Omega_1(w)&\Omega_1 (w_1)&...&\Omega_1(w_{M_{\rm tot}})\cr
...&...&...&...\cr
\Omega_{M_{\rm tot}} (w)&\Omega_{M_{\rm tot}} (w_1)
&...&\Omega_{M_{\rm tot}} (w_{M_{\rm tot}}) }
\left( \det \Omega_L(w_j)\right)^{-1}. }

\vskip 1cm
\newsec{THE 4-POINT CORRELATOR}
\vskip 1cm

The operator formalism allows also the calculation of higher order
correlation functions. In this Chapter
the computation of the four-point correlator

\eqn\cfour{ <0| \gamma (z_1) \gamma (z_2) \beta (w_1) \beta (w_2)
\prod_{s=1}^{M_{\rm tot}}
\delta (\beta (u_s)) |0>. }

is presented. The Wick theorem implies

$$<0| \gamma (z_1) \gamma (z_2) \beta (w_1) \beta (w_2)
\prod_{s=1}^{M_{\rm tot}}
\exp \left( i p_s\beta (u_s)\right) |0> =$$
$$= \big( K_{\lambda} (z_1, w_1)K_{\lambda} (z_2, w_2) +
K_{\lambda} (z_1, w_2)K_{\lambda} (z_2, w_1)\big)
<0| \prod_{s=1}^{M_{\rm tot}} \exp \left( ip_s \beta (u_s) \right) |0> +$$
$$+ K_{\lambda} (z_1, w_1) \sum_{t=1}^{M_{\rm tot}} ip_t K_{\lambda}
(z_2,u_t)
<0| \beta (w_2) \prod_{s=1}^{M_{\rm tot}} \exp \left( ip_s
\beta (u_s) \right) |0>$$
$$+ K_{\lambda} (z_1, w_2) \sum_{t=1}^{M_{\rm tot}} ip_t K_{\lambda}
(z_2, u_t)
<0| \beta (w_1) \prod_{s=1}^{M_{\rm tot}} \exp \left( ip_s
\beta (u_s) \right) |0>$$
$$+ K_{\lambda} (z_2, w_1) \sum_{t=1}^{M_{\rm tot}} ip_t K_{\lambda}
(z_1, u_t)
<0| \beta (w_2) \prod_{s=1}^{M_{\rm tot}} \exp \left( ip_s
\beta (u_s) \right) |0>$$
$$+ K_{\lambda} (z_2, w_2) \sum_{t=1}^{M_{\rm tot}} ip_t K_{\lambda}
(z_1, u_t)
<0| \beta (w_1) \prod_{s=1}^{M_{\rm tot}} \exp \left( ip_s
\beta (u_s) \right) |0>$$
\eqn\cfora{ -\sum_{t_1,t_2=1}^{M_{\rm tot}} p_{t_1}p_{t_2} K_{\lambda}
(z_1, u_{t_1})
K_{\lambda} (z_2, u_{t_2})
<0| \beta (w_1)
\beta (w_2) \prod_{s=1}^{M_{\rm tot}} \exp \left( ip_s
\beta (u_s) \right) |0> }

Only the last term in \cfora\ requires a new analysis.

It is necessary to compute

$$\beta (w_1) \beta (w_2)
\exp \left( i\sum_{s=0}^{N-1}\sum_{j=-\infty}^{+\infty} A_{s,j}\beta_{s,j}
\right) |0> =$$

$$= -{1\over 2}\sum_{r,s=0}^{N-1}\sum_{j,k=-\infty}^{+\infty}
\Psi_{r,-j}\Psi_{s,-k}
\left( x_{r,j} -i {\partial\over \partial x_{r,j}}\right)
\left( x_{s,k} -i {\partial\over \partial x_{s,k}}\right)\times$$

$$\times\exp \left( - {i\over 2}\sum_{s'=0}^{N-1}\sum_{n'<\lambda}
x_{s',n'}^2\right)
\exp \left( -{i\over 2} \sum_{q=0}^{N-1}\sum_{m\geq\lambda}
\left( (A_{q,-m} +i\sqrt{2}x_{q,m})^2 + x_{q,m}^2\right) \right) =$$

$$= \sum_{r,s=0}^{N-1}\sum_{j,k\geq\lambda}
\Psi_{r,-j}\Psi_{s,-k}
\left( (A_{r,-j} +i\sqrt{2}x_{r,j})(A_{s,-k} +i\sqrt{2}x_{s,k})
+i\delta_{jk}\right)\times$$

\eqn\danb{
\times\exp \left( - {i\over 2}\sum_{s'=0}^{N-1}\sum_{n'<\lambda}
x_{s',n'}^2\right)
\exp \left(
-{i\over 2} \sum_{q=0}^{N-1}\sum_{m\geq\lambda}
\left( (A_{q,m} +i\sqrt{2}x_{q,-m})^2 + x_{q,-m}^2\right) \right) }

where

\eqn\ufo{\beta (w) =\sum_{s=0}^{N-1}
\sum_{j=-\infty}^{+\infty}\Psi_{s,j}\beta_{s,j},\qquad
\Psi_{s,j}(w) = f_s (w) w^{-j-\lambda}
}

so that

\eqn\ufp{
A_{r,j} = \sum_{s=1}^{M_{\rm tot}} p_s \Psi_{r,j} (u_s).
}

In the correlator
$$<0|\beta (w_1)\beta (w_2) \exp \left( i\sum_{r=1}^{M_{\rm
tot}}p_r\beta (u_r)\right) |0>=$$

$$= {1\over {\bf V_{\lambda}}}\int
\prod_{s=0}^{N-1} \prod_{j=-\infty}^{+\infty}dx_{s,j} \int
\prod_{r=1}^{M_{\rm tot}}dp_r
\exp\left( {i\over 2}\sum_{t=0}^{N-1}\big( \sum_{n\geq\lambda +M_s}
x_{s,n}^2 - \sum_{n<\lambda + M_s} x_{s,n}^2\big)\right)\times$$

$$\times\sum_{r,s=1}^{N-1}\sum_{j,k\geq\lambda} \Psi_{r,-j}\Psi_{s,-k}
\left( (A_{r,-j} +i\sqrt{2}x_{r,j})(A_{s,-k} + i\sqrt{2}x_{s,k})
+i\delta_{j,k}\delta_{r,s}\right)\times$$

\eqn\ufq{
\times\exp\left( -{i\over 2} \sum_{s=0}^{N-1}\left(
\sum_{n<\lambda}x_{s,n}^2 + \sum_{j\geq\lambda}\left( (A_{s,-j}
+i\sqrt{2}x_{s,j})^2 + x_{s,j}^2\right)\right)\right)
}

the integrations over $dx_{s,n}$ for $n<\lambda$ are
trivial. In the integrations over $dx_{s,n}$ for $n\geq\lambda
+M_s$ a crucial observation is that in the sum over $j,k>\lambda$
the only terms that
survive are those for which $j,k\leq\lambda +M_s-1$.
The reason is that \nf\ holds.
One obtains therefore

$$<0|\beta (w_1)\beta (w_2) \exp \left( i\sum_{r=1}^{M_{\rm
tot}}p_r\beta (u_r)\right)=
{1\over {\bf V'_{\lambda}}} \int
\prod_{s=0}^{N-1}\prod_{n=\lambda}^{\lambda +M_s-1}dx_{s,n}
\int \prod_{r=1}^{M_{\rm tot}} dp_r\times$$

$$\times\sum_{r,s=1}^{N-1}\sum_{j,k=\lambda}^{\lambda +M_s-1}
\Psi_{r,-j}\Psi_{s,-k}
\left( (A_{r,-j} +i\sqrt{2}x_{r,j})(A_{s,-k} + i\sqrt{2}x_{s,k})
+i\delta_{j,k}\delta_{r,s}\right)\times$$

\eqn\dane{
\times\exp\left( -{i\over 2}\sum_{s=0}^{N-1}\sum_{n=\lambda}^{\lambda
+M_s-1}\left( (A_{s,-n}+i\sqrt{2}x_{s,n})^2 + 2
x_{s,n}^2\right)\right).
}

At this point it
is very convenient to rewrite above expression using the indices $L$
introduced in Chapter 3.

$$<0|\beta (w_1)\beta (w_2) \exp \left( i\sum_{r=1}^{M_{\rm
tot}}p_r\beta (u_r)\right) |0>=
{1\over {\bf V'_{\lambda}}} \int
\prod_{J=1}^{M_{\rm tot}}
\int \prod_{r=1}^{M_{\rm tot}}dp_r\times$$

$$\times\sum_{K,L=1}^{M_{\rm tot}} \Psi_K\Psi_L
\left( (A_K +i\sqrt{2}x_K)(A_L + i\sqrt{2}x_L)
+i\delta_{K,L}\right)\times$$
\eqn\danf{
\times\exp\left( -{i\over 2}\sum_{J=1}^{M_{\rm tot}}
\left( (A_J+i\sqrt{2}x_J^2 + 2
x_J^2\right)
\right).
}

One should not forget that in \cfora\ this expression is multiplied by
$(-)p_{t_1}p_{t_2}$. Performing the
 standard changes of variables $p_r\rightarrow A_L$
and $A_J\rightarrow \tilde A_J = A_J +i\sqrt{2}x_J$, the
remaining integrals are

$${1\over {\bf V'_{\lambda}}} \int
\prod_{K=1}^{M_{\rm tot}} dx_K
\int \prod_{J=1}^{M_{\rm tot}} d\tilde A_J
\exp\left(
-{i\over 2}\sum_{J=1}^{M_{\rm tot}}
\left( \tilde A_J^2 + 2 x_J^2\right)\right)\times
$$

\eqn\dang{
\times R^{-1}_{t_1M}R^{-1}_{t_2N}(\tilde A_M -i\sqrt{2}x_M)(\tilde A_N
-i\sqrt{2}x_N) \sum_{K,L=1}^{M_{\rm tot}} \Psi_K\Psi_L
\left( \tilde A_K \tilde A_L +i\delta_{K,L}\right)
}

One can replace $\Psi_J (z)$ with $\Omega_J(z)$
as for the allowed
values of $J$ only the zero modes
appear in the above expression, which we denote with the symbol
$\Omega_J$.

Integrals to be evaluated are

\eqn\danh{
\int\prod_{Q=1}^{M_{\rm tot}} d\tilde A_Q
(\tilde A_M -i\sqrt{2}x_M)(\tilde A_N -i\sqrt{2}x_N)(\tilde A_K\tilde
A_L + i\delta_{K,L})\exp \left(-{i\over 2}\sum_{S=1}^{M_{\rm tot}}\tilde
A_S^2\right)
}

It is clear that terms with $x_M$ drop out. Terms linear in $x_M$ give
rise to terms linear or of the third order in $\tilde A_P$ which yield
no contribution after integration.
Also the quadratic term containing $x_M x_N$ gives rise to integral
with $\tilde A_K\tilde A_L +i\delta_{K,L}$ which integrated give $0$
because of \nf\ .

The tensor structure of \danh\ implies that it must be of the
form

$$const \left(
\delta_{N,K}\delta_{M,L}
+\delta_{N,L}\delta_{M,K}\right)
\int\prod_{S=1}^{M_{\rm tot}}d\tilde
A_S \exp
\left( -{i\over 2} \sum_{S=1}^{M_{\rm tot}}\tilde A_S^2\right) .$$

Using

\eqn\apce{ \int dp p^4 \exp \big(-{i\over 2} p^2\big)
= -3 \int dp \exp \big(-{i\over 2} p^2 \big)}

and

\eqn\apce{ \int dp p^2 \exp \big(-{i\over 2} p^2\big)
= -i \int dp \exp \big(-{i\over 2} p^2\big) . }

it is easy to verify that $const = -1$.

After performing the
remaining trivial integrations the final expression for the
last term in \cfora\ is (apart from the Jacobian $\det \left( {\partial A_J
\over p_r}\right)$)

\eqn\apcg{
\sum_{t_1,t_2=1}^{M_{\rm tot}}
K_{\lambda} (z_1, u_{t_1}) K_{\lambda} (z_2, u_{t_2})
\Big( R_{t_1L}^{-1}\Omega_L(w_2)R_{t_2K}^{-1}\Omega_K(w_1)
+ R_{t_1K}^{-1}\Omega_K(w_1)R_{t_2L}^{-1}\Omega_L(w_2)\Big) }

Altogether \cfora\ is equal to

$${1\over \det \Omega_I(u_J)} \Bigg(
K_{\lambda} (z_1, w_1) K_{\lambda} (z_2, w_2) +
K_{\lambda} (z_1, w_2) K_{\lambda} (z_2, w_1) +$$
$$- \sum_{t=1}^{M_{\rm tot}} \sum_{K=1}^{M_{\rm tot}}\Big(
K_{\lambda} (z_1, w_1)K_{\lambda} (z_2, u_t)R_{tK}^{-1}\Omega_K (z_2)
+ K_{\lambda} (z_1, w_2)K_{\lambda} (u_t, w_1)R_{tK}^{-1}\Omega_K (z_2) +$$
$$+ K_{\lambda} (z_2, w_1)K_{\lambda} (u_t, w_2)R_{tK}^{-1}\Omega_K (z_1)
+ K_{\lambda} (z_1, w_1)K_{\lambda} (u_t, w_2)R_{tK}^{-1}\Omega_K
(z_1)\Big) +$$
$$+ \sum_{t_1,t_2=1}^{M_{\rm tot}}\sum_{L,K=1}^{M_{\rm tot}}
\Big( K_{\lambda} (u_{t_1}, w_1)K_{\lambda} (u_{t_2}, w_2)
R_{t_1L}^{-1}\Omega_L (z_2) R_{t_2K}^{-1}\Omega_K (z_1) +$$
\eqn\apch{
+ K_{\lambda} (u_{t_1}, w_1)K_{\lambda} (u_{t_2}, w_2)
R_{t_1K}^{-1}\Omega_K (z_1) R_{t_2L}^{-1}\Omega_L (z_2) \Big) \Bigg) }

With manipulations analogous to those used in the derivation of \tftb\
it is possible to write down \apch\ in a compact form. The
details are given in
Appendix B. The four-point correlator is

$${<0|\beta (z_1)\beta (z_2) \gamma (w_1) \gamma (w_2) \prod_{s=1}^{M_{\rm tot}} \delta
(\beta (u_s)) |0>\over
<0|\prod_{s=1}^{M_{\rm tot}} \delta (\beta (u_s)) |0>} = $$

$$= {<0|\beta (z_1)\gamma (w_1) \prod_{s=1}^{M_{\rm tot}} \delta
(\beta (u_s)) |0>\over
<0|\prod_{s=1}^{M_{\rm tot}} \delta (\beta (u_s)) |0>}
{<0|\beta (z_2)\gamma (w_2) \prod_{s=1}^{M_{\rm tot}} \delta
(\beta (u_s)) |0>\over
<0|\prod_{s=1}^{M_{\rm tot}} \delta (\beta (u_s)) |0>} +$$
\eqn\apcj{ +
{<0|\beta (z_1)\gamma (w_2) \prod_{s=1}^{M_{\rm tot}} \delta
(\beta (u_s)) |0>\over
<0|\prod_{s=1}^{M_{\rm tot}} \delta (\beta (u_s)) |0>}
{<0|\beta (z_2)\gamma (w_1) \prod_{s=1}^{M_{\rm tot}} \delta
(\beta (u_s)) |0>\over
<0|\prod_{s=1}^{M_{\rm tot}} \delta (\beta (u_s)) |0>}
}

\vskip 1cm
\appendix{A}{GAUSSIAN REPRESENTATION}
\vskip 1cm

The aim of this Appendix is to provide a short but
selfconsistent presentation of the
construction of "gaussian" representation for $\beta -\gamma$ system
proposed in \ag . The notation correspond to that used in Chapter 2 where
the RS is a complex sphere.

The elementary excitations for $\beta$ and $\gamma$ satisfy the
commutation
relations:

\eqn\acr{[\gamma_n, \beta_m] = \delta_{n+m,0}.
}

They are represented as

\eqn\brep{\beta_n = - {i\over \sqrt{2}} (x_{-n} -i {\partial \over
\partial x_{-n}})}

\eqn\grep{\gamma_n =  {1\over \sqrt{2}} (x_{n} +i {\partial \over
\partial x_{n}})}

The operators $\beta_n$ and $\gamma_m$ act on functions
of $x_n$. The vector space they act on is a product of
"gaussian" factors
$\exp (\pm {i\over 2}x_n^2)$ multiplied by polynomials in $x_m$.
In this vector "Fock space" the scalar product is defined in the
following way:

\eqn\sprep{ <\Phi_1 |\Phi_2> \equiv {1\over V_{\lambda}}
\int\prod_n dx_n \left(
\Phi_1(x)\right)^*\Phi_2(x). }

It is assumed that $x_n^*= x_{-n}$ (but in the case of general RS
this rule will have to be modified).
$V_{\lambda}$ is a formally infinite normalization factor equal to

\eqn\apvnorm{
V_{\lambda}= \Big( {\int dp\exp \big( -{i\over 2} p^2\big)
\over 2\pi}\Big)^{2\lambda -1}
\Big(\prod_{m<-\lambda} \int \exp \left( -i x_m^2\right) dx_m\Big)
\Big(\prod_{n\geq 1-\lambda}\int \exp \left( i x_n^2\right) dx_n\Big)
}

It is understood that \sprep\ is defined in such a way that the two formally
infinite terms in numerator and denominator
cancel each other.

A choice of the vaccum state is a choice of "Bose sea level".
A vacuum state
$\Phi^{(\lambda )}$ can be constructed such that

\eqn\vacdefa{ <\Phi^{(\lambda )}| \Phi^{(\lambda )}> = \infty }

(the explanation for that infinity can be found in Introduction)
but (assume that $\lambda\geq 2$)

\eqn\vacdefb{ <\Phi^{(\lambda )}|\prod_{j=1-\lambda}^{\lambda -1}
\delta (\gamma_j )
|\Phi^{(\lambda )}> = 1. }

The correct defintion is

\eqn\vacdefc{ \Phi^{\lambda}(x) =  \exp {i\over 2} \left(
\sum_{n\geq \lambda} x_n^2 - \sum_{n<\lambda}x_n^2\right). }

It satisfies

\eqn\vacdefd{ \left( \Phi^{({\lambda})} (x)\right)^*
= \Phi^{1-\lambda } (x). }

so that

\eqn\vacdefe{ \Phi_0^{(\lambda )}\left( \Phi_0^{(\lambda )}\right)^* =
\exp i \left( \sum_{n\geq\lambda}x_n^2 - \sum_{n<-\lambda} x_n^2
\right) }

The integration over $dx_{\lambda -1} ... dx_{1-\lambda}$ in \sprep\ lead to
\vacdefa .

On the other hand

$$\delta (\gamma_n) \exp \Big( -{i\over 2} x_n^2 \Big) =
{1\over 2\pi}\int dp \exp \Big( ip\gamma_n\Big)
\exp \Big( -{i\over 2} x_n^2 \Big) =$$

$$={1\over 2\pi}
\int dp \exp \Big( {i\over \sqrt{2}}
p(x_n +i{\partial\over\partial x_n}) \Big)
\exp \Big( -{i\over 2} x_n^2\Big)
= {1\over 2\pi} \int dp \exp \Big( {i\over 2}x_n^2 -{i\over
2}(p-i\sqrt{2}x_n)^2 \Big) =$$

\eqn\vacdeff{
= {1\over 2\pi }\int dp \exp \Big( -{1\over 2}p^2\big)
\exp \left( {i\over 2} x_n^2.
\right)
}

Therefore, thanks to the
appropriate definition of the normalization factor $V_{\lambda}$, 
\vacdefb\ holds.

An important identity used above is

\eqn\tsb{ \exp \left( a{\partial\over\partial x_{-n}} \right) \exp \left(
{i\over 2}x_{-n}^2\right) 
= \exp \left( iax_{-n} + {i\over 2} a^2\right) \exp \left( 
{i\over 2}x_{-n}^2\right). }

\vskip 1cm
\appendix{B}{DERIVATION OF \apcj }
\vskip 1cm
The aim of this Appendix is to provide a proof that the
4-point correlation
function can be represented by \apcj . The starting point is \apch\ and
the basic trick is that used in the derivation of \tftb :

$$\sum_{s,N=1}^{M_{\rm tot}} K_{\lambda}(z, w_{s})
R_{sN}^{-1} \Omega_{N}(w) = K_{\lambda} (z,w)-$$

\eqn\pepa{
\det\pmatrix{
K_{\lambda}(z,w)&K_{\lambda}(z,w_1)&...&K_{\lambda}(z,w_{M_{\rm tot}})\cr
\Omega_1(w)&\Omega_1 (w_1)&...&\Omega_1(w_{M_{\rm tot}})\cr
...&...&...&...\cr
\Omega_{M_{\rm tot}} (w)&\Omega_{M_{\rm tot}} (w_1)
&...&\Omega_{M_{\rm tot}} (w_{M_{\rm tot}})}
\big(\det \Omega_L(w_j)\big)^{-1}. }

Thanks to \pepa\ all the summations in \apch\ can be eliminated. One
obtains (everything has to be multiplied by ${1\over \det \Omega_I(u_J)}$)

$$ K_{\lambda} (z_1, w_1) K_{\lambda} (z_2, w_2) +
K_{\lambda} (z_1, w_2) K_{\lambda} (z_2, w_1) +$$
$$- K_{\lambda} (z_1, w_1)K_{\lambda} (z_2, w_2)+$$
$$+ {1\over \big( \det \Omega_I(u_J)\big)}
K_{\lambda} (z_1, w_1) \det\pmatrix{
K_{\lambda}(z_2,w_2)&K_{\lambda}(z_2,u_1)&...
&K_{\lambda}(z_2,u_{M_{\rm tot}})\cr
\Omega_1(w_2)&\Omega_1 (u_1)&...&\Omega_1(u_{M_{\rm tot}})\cr
...&...&...&...\cr
\Omega_{M_{\rm tot}} (w_2)&\Omega_{M_{\rm tot}} (u_1)
&...&\Omega_{M_{\rm tot}} (u_{M_{\rm tot}})}+$$
$$-K_{\lambda} (z_1, w_2)K_{\lambda} (z_2, w_1)+$$
$$+ {1\over \big( \det \Omega_I(u_J)\big)}
K_{\lambda} (z_1, w_2) \det\pmatrix{
K_{\lambda}(z_2,w_1)&K_{\lambda}(z_2,u_1)&...
&K_{\lambda}(z_2,u_{M_{\rm tot}})\cr
\Omega_1(w_1)&\Omega_1 (u_1)&...&\Omega_1(u_{M_{\rm tot}})\cr
...&...&...&...\cr
\Omega_{M_{\rm tot}} (w_1)&\Omega_{M_{\rm tot}} (u_1)
&...&\Omega_{M_{\rm tot}} (u_{M_{\rm tot}})}+$$
$$-
K_{\lambda} (z_2, w_1)K_{\lambda} (z_1, w_2)+$$
$$+ {1\over \big( \det \Omega_I(u_J)\big)}
K_{\lambda} (z_2, w_1) \det\pmatrix{
K_{\lambda}(z_1,w_2)&K_{\lambda}(z_1,u_1)&...
&K_{\lambda}(z_1,u_{M_{\rm tot}})\cr
\Omega_1(w_2)&\Omega_1 (u_1)&...&\Omega_1(u_{M_{\rm tot}})\cr
...&...&...&...\cr
\Omega_{M_{\rm tot}} (w_2)&\Omega_{M_{\rm tot}} (u_1)
&...&\Omega_{M_{\rm tot}} (u_{M_{\rm tot}})}+$$
$$-
K_{\lambda} (z_2, w_2)K_{\lambda} (z_1, w_1)+$$
$$+ {1\over \big( \det \Omega_I(u_J)\big)}
K_{\lambda} (z_2, w_2) \det\pmatrix{
K_{\lambda}(z_1,w_1)&K_{\lambda}(z_1,u_1)&...
&K_{\lambda}(z_1,u_{M_{\rm tot}})\cr
\Omega_1(w_1)&\Omega_1 (u_1)&...&\Omega_1(u_{M_{\rm tot}})\cr
...&...&...&...\cr
\Omega_{M_{\rm tot}} (w_1)&\Omega_{M_{\rm tot}} (u_1)
&...&\Omega_{M_{\rm tot}} (u_{M_{\rm tot}})}+$$
$$
+ \Bigg(
K_{\lambda} (z_1, w_2)
-{1\over \big( \det \Omega_I(u_J)\big)}
\det\pmatrix{
K_{\lambda}(z_1,w_2)&K_{\lambda}(z_1,u_1)&...
&K_{\lambda}(z_1,u_{M_{\rm tot}})\cr
\Omega_1(w_2)&\Omega_1 (u_1)&...&\Omega_1(u_{M_{\rm tot}})\cr
...&...&...&...\cr
\Omega_{M_{\rm tot}} (w_2)&\Omega_{M_{\rm tot}} (u_1)
&...&\Omega_{M_{\rm tot}} (u_{M_{\rm tot}})}\Bigg)\times$$
$$
\times\Bigg(
K_{\lambda} (z_2, w_1)
- {1\over \big( \det \Omega_I(u_J)\big)}
\det\pmatrix{
K_{\lambda}(z_2,w_1)&K_{\lambda}(z_2,u_1)&...
&K_{\lambda}(z_2,u_{M_{\rm tot}})\cr
\Omega_1(w_1)&\Omega_1 (u_1)&...&\Omega_1(u_{M_{\rm tot}})\cr
...&...&...&...\cr
\Omega_{M_{\rm tot}} (w_1)&\Omega_{M_{\rm tot}} (u_1)
&...&\Omega_{M_{\rm tot}} (u_{M_{\rm tot}})}\Bigg) +$$
$$+ \Bigg( K_{\lambda} (z_1, w_1)
- {1\over \big( \det \Omega_I(u_J)\big)}
\det\pmatrix{
K_{\lambda}(z_1,w_1)&K_{\lambda}(z_1,u_1)&...
&K_{\lambda}(z_1,u_{M_{\rm tot}})\cr
\Omega_1(w_1)&\Omega_1 (u_1)&...&\Omega_1(u_{M_{\rm tot}})\cr
...&...&...&...\cr
\Omega_{M_{\rm tot}} (w_2)&\Omega_{M_{\rm tot}} (u_1)
&...&\Omega_{M_{\rm tot}} (u_{M_{\rm tot}})}\Bigg)\times$$
\eqn\pepo{
\times\Bigg( K_{\lambda} (z_2, w_2)
- {1\over \big( \det \Omega_I(u_J)\big)}
\det\pmatrix{
K_{\lambda}(z_2,w_2)&K_{\lambda}(z_2,u_1)&...
&K_{\lambda}(z_2,u_{M_{\rm tot}})\cr
\Omega_1(w_2)&\Omega_1 (u_1)&...&\Omega_1(u_{M_{\rm tot}})\cr
...&...&...&...\cr
\Omega_{M_{\rm tot}} (w_2)&\Omega_{M_{\rm tot}} (u_1)
&...&\Omega_{M_{\rm tot}} (u_{M_{\rm tot}})}\Bigg)}

All except of two terms cancel each other
and \apch\ turns out to be equal

$$
{1\over \big(\det \Omega_I(u_J)\big)^3}
\Bigg(
\det\pmatrix{
K_{\lambda}(z_1,w_2)&K_{\lambda}(z_1,u_1)&...
&K_{\lambda}(z_1,u_{M_{\rm tot}})\cr
\Omega_1(w_2)&\Omega_1 (u_1)&...&\Omega_1(u_{M_{\rm tot}})\cr
...&...&...&...\cr
\Omega_{M_{\rm tot}} (w_2)&\Omega_{M_{\rm tot}} (u_1)
&...&\Omega_{M_{\rm tot}} (u_{M_{\rm tot}})}\Bigg)\times$$
$$\times\Bigg(
\det\pmatrix{
K_{\lambda}(z_2,w_1)&K_{\lambda}(z_2,u_1)&...
&K_{\lambda}(z_2,u_{M_{\rm tot}})\cr
\Omega_1(w_1)&\Omega_1 (u_1)&...&\Omega_1(u_{M_{\rm tot}})\cr
...&...&...&...\cr
\Omega_{M_{\rm tot}} (w_1)&\Omega_{M_{\rm tot}} (u_1)
&...&\Omega_{M_{\rm tot}} (u_{M_{\rm tot}})}\Bigg) +$$

$$+ \Bigg(
\det\pmatrix{
K_{\lambda}(z_1,w_1)&K_{\lambda}(z_1,u_1)&...
&K_{\lambda}(z_1,u_{M_{\rm tot}})\cr
\Omega_1(w_1)&\Omega_1 (u_1)&...&\Omega_1(u_{M_{\rm tot}})\cr
...&...&...&...\cr
\Omega_{M_{\rm tot}} (w_2)&\Omega_{M_{\rm tot}} (u_1)
&...&\Omega_{M_{\rm tot}} (u_{M_{\rm tot}})
}\Bigg)\times$$
\eqn\pepsi{
\times\Bigg(
\det\pmatrix{
K_{\lambda}(z_2,w_2)&K_{\lambda}(z_2,u_1)&...
&K_{\lambda}(z_2,u_{M_{\rm tot}})\cr
\Omega_1(w_2)&\Omega_1 (u_1)&...&\Omega_1(u_{M_{\rm tot}})\cr
...&...&...&...\cr
\Omega_{M_{\rm tot}} (w_2)&\Omega_{M_{\rm tot}} (u_1)
&...&\Omega_{M_{\rm tot}} (u_{M_{\rm tot}})}\Bigg)\Bigg)
}

By looking at \tfina\ one immediately gets \apcj .

\listrefs
\bye